# Quasi-Single-Mode Transmission over Ultra-Low-Loss Few-Mode Fibre for Data Centre Interconnects

Fabio A. Barbosa, *Member, IEEE*, Rostislav R. Khrapko, Ming-Jun Li, *Fellow, IEEE*, Filipe M. Ferreira*, Senior Member, IEEE*

***Abstract*—A novel ultra-low-loss few-mode fibre is experimentally investigated for quasi-single-mode transmission over data centre interconnect distances. Multipath interference is characterised across the C-band matching a modified Gaussian noise model to experimental results obtained with conventional training-sequence-based digital signal processing, enabling in-service assessment without traffic disruption. This model shows strong agreement with power-fluctuation measurements performed on a dedicated unmodulated-laser setup. The impact of equaliser length and digital backpropagation on system performance is also analysed. Compared with standard ultra-low-loss single-mode fibres, transmission experiments reveal only minor signal-to-noise ratio degradation for up to 42-Gbaud DP-256-QAM over a 24-km span. These findings highlight the potential of ultra-low-loss few-mode fibre for scalable, high-capacity data centre interconnect applications while offering potential for future space-division multiplexing upgrades.**

## I. INTRODUCTION

Data centre networks have witnessed a rapid surge in throughput demand, primarily driven by the expansion of cloud computing and artificial intelligence applications. To accommodate this growth, scalability and high integration density have become critical [1, 2], with space-division multiplexing (SDM) emerging as a potential solution. Considerable efforts have been directed toward introducing scalable data centre architectures. Recently, optical circuit switching (OCS)-based designs have enabled flexible transceiver upgrades while retaining existing cabling infrastructure [3]. Building on these developments, SDM fibres engineered for single-mode transmission yet inherently supporting spatial diversity, presents a compelling opportunity for future capacity upgrades, while maintaining compatibility with current network infrastructure.

Quasi-single-mode (QSM) transmission, characterised by targeting the transmission of only the $LP_{01}$ mode in few-mode fibres (FMFs), has been investigated experimentally mostly in the long-haul context. In [4-6], this concept was used to explore the large effective area of FMFs and increase transmission reach at the presence of reduced nonlinear interference (NLI). The lower NLI benefit, however, typically comes at the cost of the linear multipath interference (MPI) impairment. Signal copies that couple to high-order modes at discrete points (e.g. launching or splices) or due to imperfections along the fibre interfere with the signal in the target mode, resulting in MPI. Long adaptive multiple-input multiple-output (MIMO) equalisers have been investigated for MPI mitigation in [4]. In [5], a hybrid fibre span including a single-mode fibre (SMF) and FMF was tested, together with multi-carrier modulation formats, to reduce MPI penalties.

Recently, a novel alkali-doped, silica-core ultra-low-loss (ULL) few-mode fibre (FMF) supporting both the $LP_{01}$ and $LP_{11}$ modes has been presented [7]. By leveraging an improved design of a larger core (19-μm diameter) and draw conditions, the $LP_{01}$ showed a measured loss of 0.1400 dB/km at 1560 nm and 0.1407 dB/km at 1550 nm for a 24-km spool, and slightly higher losses for the $LP_{11}$. Further analysis observed that power coupling to $LP_{11}$ was not responsible for a significant attenuation increase.

This study investigates QSM transmission over the novel ULL FMF targeting data centre interconnect (DCI) applications. A standard digital signal processing (DSP) chain is employed, and the ULL FMF overall performance benchmarked against state-of-the-art SMF. This work extends the results in [8], assessing different equalisation approaches and the impact of equaliser length on system performance. The new analysis also considers digital backpropagation (DBP). MPI is quantified across C-band using different methods, including a modified Gaussian noise model that enables in-service assessment without traffic disruption. The performance results show small signal-to-noise ratio (SNR) penalties for transmission of up to 42-Gbaud DP-256-QAM, highlighting the ULL FMF potential for DCI applications.

## II. ESTIMATION OF MULTIPATH INTERFERENCE

QSM transmission is subject to MPI. In this work, MPI is estimated with two approaches. The 1st MPI estimation method uses data traces acquired with a coherent receiver. As discussed in [9], the total received SNR after coherent detection, $SNR_{Total}$, can be modelled as the combination of

Received xx xx xxxx; revised xx xx xxxx; accepted xx xx xx. Date of publication x xx xxxx; date of current version xx xx xxxx. This work was supported in part by the UKRI Future Leaders Fellowship under the Grant MR/Y034260/1. *(Corresponding author: Fabio A. Barbosa)*.

Fabio A. Barbosa and Filipe M. Ferreira are with the Optical Networks Group, University College London (UCL), WC1E 6BT London, U.K. (e-mail: fabio.barbosa@ucl.ac.uk).

Rostislav R. Khrapko is with the Department of Optical Fiber Research and Development, Corning Incorporated, Corning, NY 14831 USA.

Ming-Jun Li is with the Science and Technology Division, Corning Incorporated, Corning, NY 14831 USA.

Color versions of one or more of the figures in this article are available online at http://ieeexplore.ieee.org

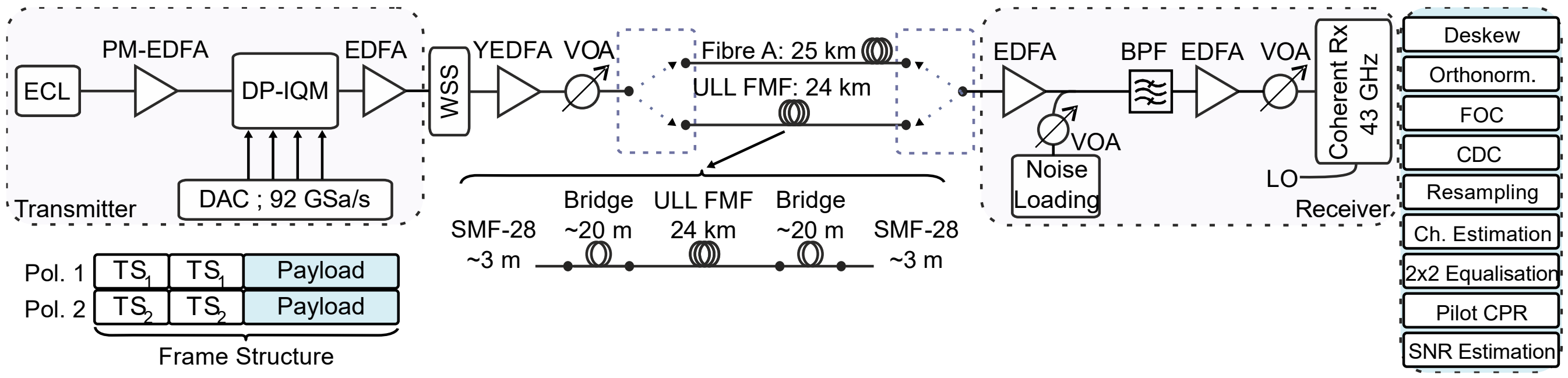

**Fig. 1** Schematic diagram of the optical transmission system used for single-channel transmission.

statistically independent components such that

$$SNR_{Total}^{-1} = SNR_{TRX}^{-1} + SNR_{ASE}^{-1} + SNR_{NLI}^{-1} + SNR_{MPI}^{-1}, \quad (1)$$

where $SNR_{TRX}$ is the transceiver-limit SNR, $SNR_{ASE}$ is given by the ASE noise and $SNR_{NLI}$ arises from fibre nonlinearities. Similarly to [10, 11], the MPI contribution is modelled as an additive noise and added to (1) as the last term.

In this work, unlike [10, 11], (1) is matched to experimental results for estimation of the multiple factors impacting system performance across all optical power regimes, including MPI at multiple wavelengths across C-band. Each component is estimated separately: $SNR_{TRX}$ is estimated in back-to-back with the highest achievable optical signal-to-noise ratio (OSNR), $SNR_{NLI}$ is estimated well within the nonlinear power regime, $SNR_{ASE}$ is estimated at a launch power within the linear regime and under limited OSNR, ensuring that the achievable SNR is much smaller than $SNR_{TRX}$. This leaves just $SNR_{MPI}$ to be matched, and it is worth noting that MPI ultimately limits performance at optimum launch power [6, 12]. Substituting in (1) the values for only $SNR_{TRX}$, $SNR_{NLI}$ and $SNR_{ASE}$ estimated from the experimental traces, one could obtain $SNR_{Total}$ excluding MPI contributions. The difference between this quantity and the experimental results for an MPI-impaired fibre is an estimate of $SNR_{MPI}$.

The 2nd method uses power measurements such that

$$MPI = 20\log_{10}(\sigma/P_{avg}), \quad (2)$$

where σ is the standard deviation of power samples and $P_{ave}$ is the average power [13].

## III. Experimental Setup

The experimental setup explored in this work is shown in Fig. 1. The optical carrier from a tuneable external cavity laser (ECL) with <100-kHz linewidth was modulated via a dual-polarisation (DP) IQ modulator driven by a 92 GSa/s arbitrary waveform generator (AWG). The drive signals were DP-256-QAM signals, shaped with a root raised cosine filter with 0.1% roll-off and linearly pre-distorted to compensate AWG and modulator frequency roll-off. Experiments included 10-Gbaud single sideband (SSB) and 42-Gbaud conventional signals. SSB transmission is employed to circumvent I/Q imbalances from both electrical and optical components [14], pushing parasitic signal images out-of-band and preventing their misidentification as MPI-impact on system performance. Ultimately, this approach increases $SNR_{TRX}$. The SSB signals were generated digitally shifting baseband 10 Gbaud signals by 6 GHz. As per the inset in Fig. 1, the frames comprised two repeated 1024-symbol constant amplitude zero autocorrelation (CAZAC) training sequences (TSs), generated using the concept of shift-orthogonality (i.e., $TS_2$ is a cyclic-shifted version of $TS_1$) [15], and a $2^{16}$-symbol payload, with 4-QAM pilot symbols at a rate of 1/32 for carrier phase recovery (CPR).

This study assessed QSM transmission over a 24-km ULL FMF supporting the $LP_{01}$ and $LP_{11}$ modes [7]. By scanning offset launching, the differential mode dispersion (DMD) was estimated to be 1.5 ns/km for the spool under investigation. Due to its larger core, the ULL FMF features a $LP_{01}$ with a mode field diameter (MFD) of 14.2 μm at 1550 nm, compared to ~10.5-μm for standard SMFs. To ensure low loss $LP_{01}$ launching, it is customary to use conditioning *bridge* fibres. At the receiver side, *bridge* fibres help filtering out high-order modes that might have been excited along the link. In this work, 20 m of a large effective area SMF with a MFD of 14 μm (~150 μm$^2$ eff. area) were used. The *bridge* fibre was spooled with 60-mm diameter seeking additional mode stripping at the receiver side. Additionally, to reduce losses at the transmitter/receiver interfaces, ~3-m SMF-28 pigtails were spliced to the *bridge* fibres. A total loss of 1.4 dB was added to ULL FMF spool stemming from splicing (0.48dB at each end) and spooling of the *bridge* fibre. Figure 1 shows a schematic of the conditioning fibres interfacing the ULL FMF. As reference, a 25-km ULL standard effective area (82 μm$^2$) SMF was used. This fibre is referred to as Fibre A.

A single-span setup was used, with launch power set by a booster ytterbium/erbium-dopped fibre amplifier (YEDFA) and a variable optical attenuator (VOA). It included a noise loading stage to set the OSNR at 38 dB for a 0 dBm launch power using Fibre A at 1550 nm, remaining unchanged throughout the experiment. Heterodyne detection employed a 43-GHz coherent receiver front-end, a <100-kHz linewidth local oscillator (LO), and an 80 GSa/s real-time oscilloscope. In the SSB case, the LO was offset by 11 GHz from the transmitter laser to avoid receiver-side I/Q imbalances [14].

For each configuration, five traces were acquired and processed off-line with the DSP chain shown in Fig. 1. Channel estimation was performed with a time-domain least-square (LS) approach [15]. The 2×2-MIMO channel impulse responses (CIRs) were truncated to *L* taps, zero-padded and converted to frequency domain for zero-forcing equalisation. An overlap-save method with 1024-point FFTs and 128-sample overlap was used. The traces were also processed by a DSP chain using a fully supervised time-domain equaliser (TDE) based on the radius directed equaliser (RDE), replacing the LS estimation. DBP (single mode) was also assessed with 10 steps for both fibres. The effective SNR of received constellations was used as performance metric. Traces were acquired from 1535 to 1565 nm in the ITU grid. All results were averaged over five collected traces. Unless stated otherwise, results use LS-estimated CIRs truncated to $L = 41$.

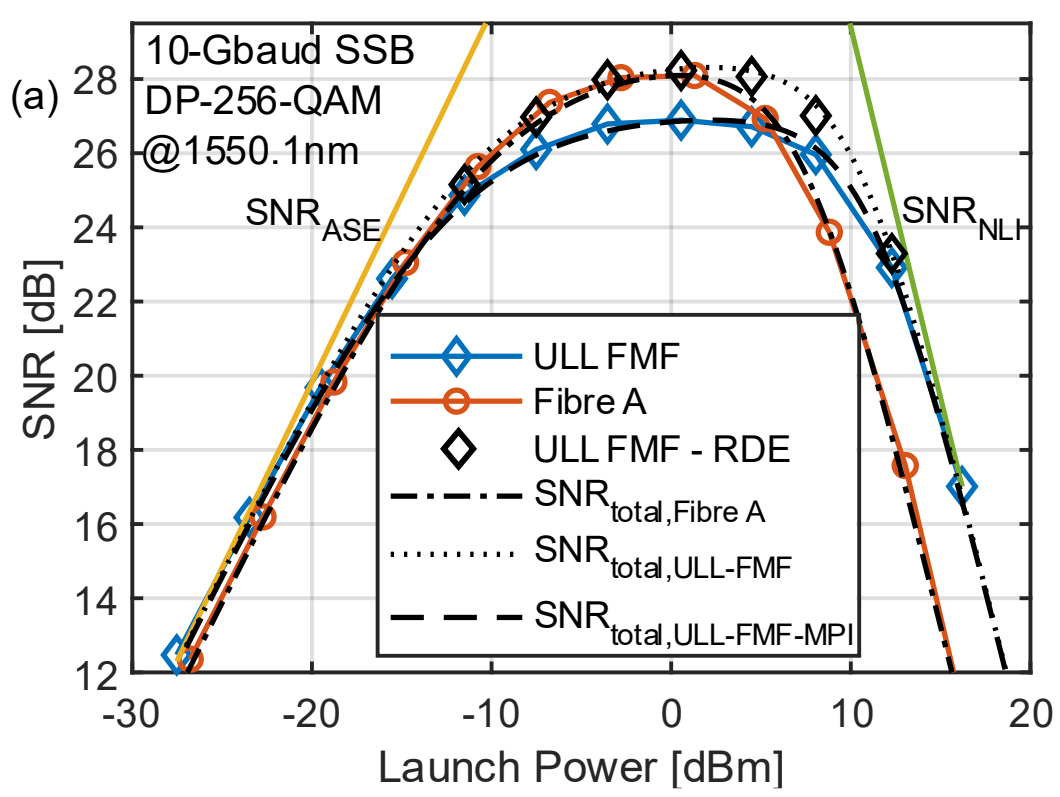

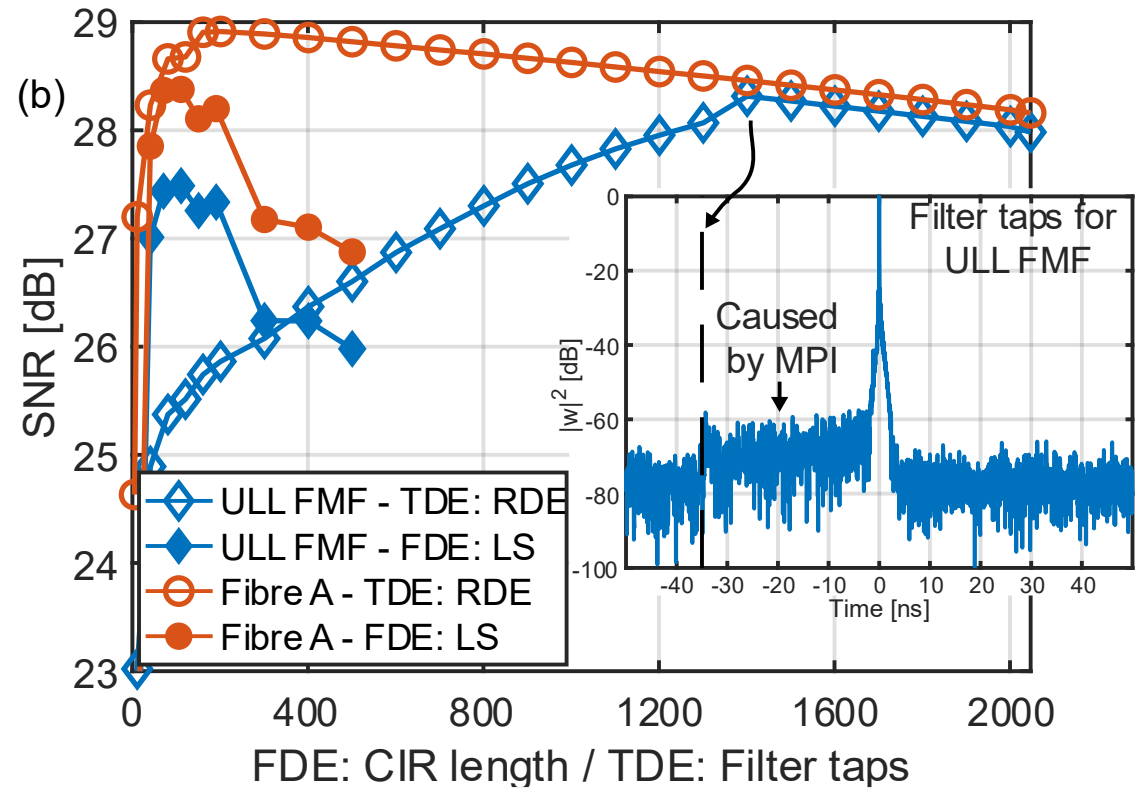

**Fig. 2** For 10-Gbaud SSB DP-256-QAM transmission over 24-km ULL FMF and 25-km Fibre A, (a) received SNR vs launch power, including model fitting, and (b) received SNR vs (FDE) CIR length and (TDE) filter taps at optimum launch power.

## IV. Results and Discussions

An initial assessment of the ULL FMF was carried out using 10-Gbaud SSB signals, with particular attention given to the impact of MPI on overall system performance. Figure 2(a) presents, for 1550.1 nm, the received SNR as a function of launch power for both Fibre A and ULL FMF. Owing to its lower attenuation, the ULL FMF showed higher received SNR (~0.5 dB) than Fibre A at the ASE-limit regime. Notably, further performance gains could be realized by minimizing the losses introduced by the conditioning fibres (see Sec. III). In the nonlinear interference (NLI)-limited regime, ULL FMF also outperforms Fibre A, which can be attributed to its larger effective area. However, at optimum launch power, a performance degradation (~1 dB) is observed due to MPI.

The MPI penalty was studied using the analytical model outlined in Sec. II. The $SNR_{Total}$ was estimated fitting each component of (1) to the experimental results separately. Figure 2(a) shows the results for ULL FMF and Fibre A as $SNR_{Total,ULL\text{-}FMF\text{-}MPI}$ (dashed line) and $SNR_{Total,Fibre\text{-}A}$ (dash-dotted lines), respectively, and individual contributions for the ULL FMF. The $SNR_{Total}$ for the ULL FMF excluding MPI contributions, $SNR_{Total,ULL\text{-}FMF}$ (dotted line), was also estimated. This was done by substituting in (1) only the respective estimated terms of $SNR_{TRX}$, $SNR_{ASE}$ and $SNR_{NLI}$ for the ULL FMF. Finally, $SNR_{MPI}$ was estimated by calculating the gap between $SNR_{Total,ULL\text{-}FMF}$ and the experimental measurements at optimum launch power. This approach yielded an MPI of -32.4 dB, which is consistent with previously reported findings for similar fibres [6].

To further investigate MPI, a fully supervised RDE algorithm was applied to the 10-Gbaud SSB signals. The resulting received SNR at optimum launch power versus the number of filter taps is shown in Fig. 2(b). Results obtained with the time-domain LS estimation method are also presented. For this scenario, the number of taps refer to the length of the truncated CIR, $L$. The inset in Fig. 2(b) shows that the MPI increases the amplitude of the taps on the leading edge of the filter over a ~35-ns window, which aligns with the estimated DMD of 1.5 ns/km for the 24-km spool under test. The high DMD is expected for this fibre given its step-index-like refractive index profile [7]. The RDE equaliser with sufficiently long filters significantly mitigated MPI, reducing the performance gap between the ULL FMF and Fibre A. As illustrated in Fig. 2(a) (diamond markers), the performance with the 1401-tap RDE closely matched the analytically estimated $SNR_{Total,ULL\text{-}FMF}$ ($SNR_{Total}$ for the ULL FMF excluding MPI). The investigations with the LS estimation approach were limited to 500 taps due to noise enhancement and artifacts in the estimated CIRs. Importantly, comparable performances were achieved with the LS approach using shorter CIR lengths. The fully supervised RDE was employed here to further study the MPI. Hereafter, only the time-domain LS approach with $L = 41$ is considered.

Figure 3 expands the MPI analysis across the C-band, comparing results for both 10-Gbaud SSB and 42-Gbaud DP-256-QAM signals. It also includes MPI estimates derived from power measurements. In this approach, a tuneable continuous-wave (CW) laser was swept from 1534.5 nm to 1565.5 nm with a resolution of 0.1 pm. Power samples were collected using an integration time of 0.2 ms and grouped into 1-nm bins. Then, MPI was estimated using (2), with each data point calculated from 200,000 samples. As shown in Fig. 3, MPI estimates via power measurements remain consistently around -33 dB across the C-band, closely matching those derived from the 10-Gbaud SSB signals using the analytical model. In contrast, estimates based on the 42-Gbaud signals using the analytical method exhibit significant overestimation and variability. This discrepancy is attributed to transceiver impairments being misinterpreted as MPI-like effects. A finite and frequency-dependent ENoB, noise from linear electrical amplifiers and I/Q imbalances are examples of sources of transceiver noise reducing the SNR margin at the optimum launch power and affecting the MPI estimation method.

Figure 4 reports the received SNR at optimum launch power across C-band without (unfilled markers) and with (filled markers) DBP. LS channel estimation was employed with $L$ = 41. The SNR degradation observed below 1545 nm is caused by the shifted gain profile of the YEDFA. For 10-Gbaud SSB, the ULL FMF exhibited a ~1 dB penalty when compared to Fibre A. Whereas, the gap narrows for 42-Gbaud signals, as this scenario is constrained by transceiver noise. Further improvements to the ULL FMF performance could be achieved by optimising the launch splice and *bridge* fibre. The relative SNR boost provided by DBP was larger (~0.1 dB) for Fibre A, as MPI limited the gains for the ULL FMF. Notably, improvements were more prominent for the SSB scenarios due to their inherent reduction in transceiver impairments [16]. Also, these gains were ultimately limited by the length of the fibres. Although MPI exhibits statistical characteristics, an analysis of its dynamics is beyond the scope of this work.

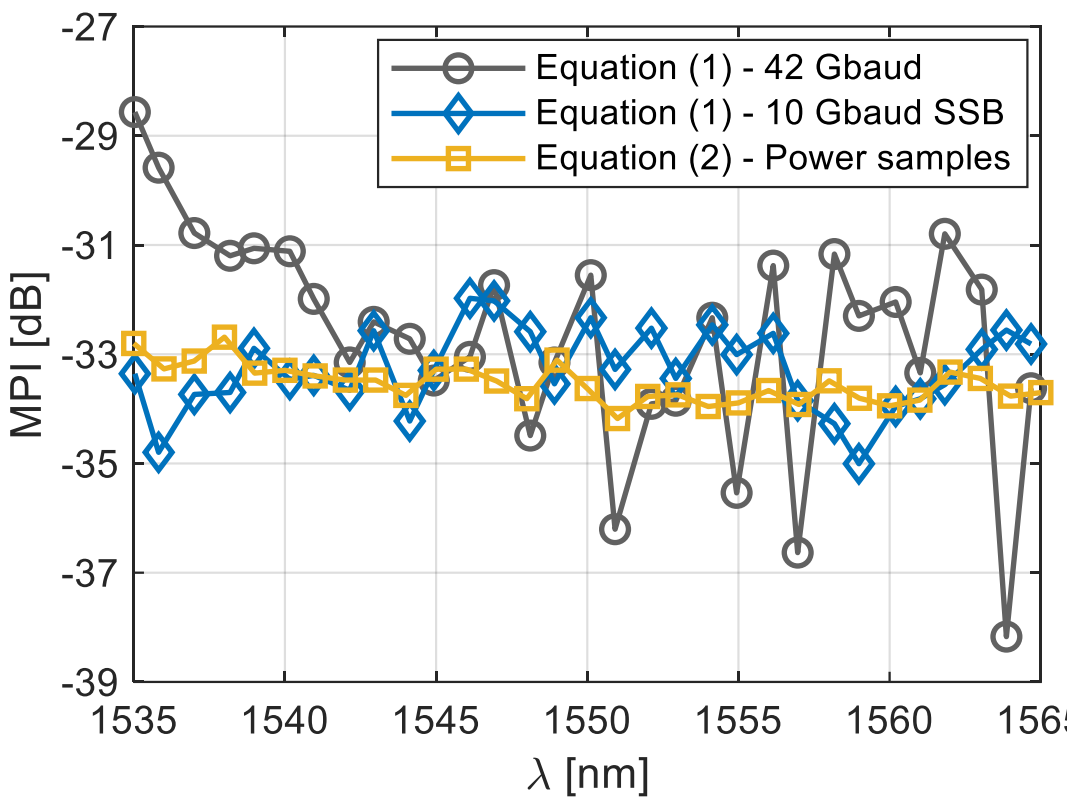

**Fig. 3** MPI vs wavelength for the ULL FMF obtained with models (1) and (2) fitted to modulated signals as well as to power samples.

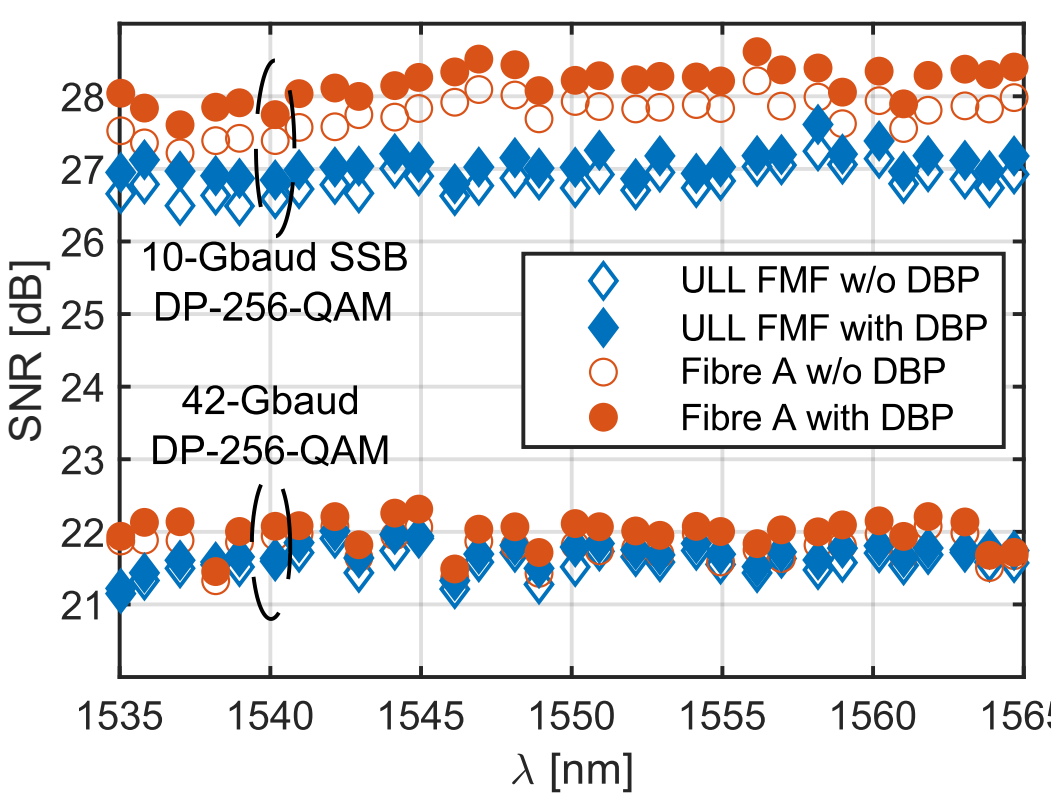

**Fig. 4** Received SNR at optimum launch power vs wavelength for ULL FMF and Fibre A obtained with 10-Gbaud SSB and 42-Gbaud signals, with and without DBP.

The results demonstrate that the ULL FMF support high-symbol rate QSM transmission with small penalties over DCI distances. In fact, the low loss resulting from its optimised design could be further explored by improving the *bridge* fibres and their splice to the ULL FMF. Importantly, the continuous development in DSP, 3D-printed photonic lanterns and wavefront shaping could unlock opportunities for future SDM upgrades [17, 18]. The SDM performance of this fibre is a topic for future investigation.

## V. Conclusion

A novel ULL FMF was experimentally investigated for QSM transmission over DCI distances. An analytical model that enables in-service MPI monitoring was investigated, showing good agreement with estimates from power samples. The ULL FMF showed small SNR penalties of ~0.3 dB using conventional DSP compared to standard SMFs for up to 42-Gbaud DP-256-QAM signals. These results reveal the potential of the ULL FMF for current DCI needs while offering potential for future SDM upgrades.